# NEURAL NETWORK AIDED GLITCH-BURST DISCRIMINATION AND GLITCH CLASSIFICATION

SALVATORE RAMPONE

*Department of Science and Technology*
*University of Sannio*
*Via dei Mulini 59/A – Palazzo Inarcassa*
*I-82100 Benevento, Italy*
*rampone@unisannio.it*

VINCENZO PIERRO*, LUIGI TROIANO[†]
and INNOCENZO M. PINTO[‡]

*Department of Engineering*
*University of Sannio*
*I-82100 Benevento, Italy*
*\*pierro@unisannio.it*
*[†]troiano@unisannio.it*
*[‡]pinto@unisannio.it*

We investigate the potential of neural-network based classifiers for discriminating gravitational wave bursts (GWBs) of a given canonical family (e.g. core-collapse supernova waveforms) from typical transient instrumental artifacts (glitches), in the data of a single detector. The further classification of glitches into typical sets is explored. In order to provide a proof of concept, we use the core-collapse supernova waveform catalog produced by H. Dimmelmeier and co-Workers, and the data base of glitches observed in laser interferometer gravitational wave observatory (LIGO) data maintained by P. Saulson and co-Workers to construct datasets of (windowed) transient waveforms (glitches and bursts) in additive (Gaussian and compound-Gaussian) noise with different signal-to-noise ratios (SNR). Principal component analysis (PCA) is next implemented for reducing data dimensionality, yielding results consistent with, and extending those in the literature. Then, a multilayer perceptron is trained by a backpropagation algorithm (MLP-BP) on a data subset, and used to classify the transients as glitch or burst. A Self-Organizing Map (SOM) architecture is finally used to classify the glitches. The glitch/burst discrimination and glitch classification abilities are gauged in terms of the related truth tables. Preliminary results suggest that the approach is effective and robust throughout the SNR range of practical interest. Perspective applications pertain both to distributed (network, multisensor) detection of GWBs, where some *intelligence* at the single node level can be introduced, and instrument diagnostics/optimization, where spurious transients can be identified, classified and hopefully traced back to their entry points.

## 1. Introduction

Gravitational waves (GWs) are ripples in the curvature of space-time which propagate as a wave, traveling outward from the source. GW astronomy is expected to open an essentially new observational window on the physical universe. Several classes of GWs of cosmic origin are currently being sought for, including continuous, transient and stochastic ones. An essential distinction among these different signals concerns our ability in modeling the expected waveforms. GW bursts (henceforth GWBs) are transient signals for which only a few physically-based models exist.[1–3] GW detectors (with specific reference to the present-day large baseline optical interferometers) are invariably affected by transient disturbances of various origins.[4] Using auxiliary channels to monitor the status of the instrument and its environment may help in identifying and vetoing these disturbances. Experimental evidence suggests that a residual impulsive component will nonetheless be present in the data. It is believed[5] that distinguishing these spurious noise glitches from true GWB of cosmic origin becomes feasible, in principle, only if the outputs of several detectors are suitably combined. In fact, various coincidence algorithms, based on consistency tests among candidate-events gathered by different detectors, have been studied and tested.[6,7] These algorithms, while conceptually simple and computationally inexpensive, turn out to be less efficient, in general, compared to coherent techniques, where the output data from several sensors are combined to form a suitable detection statistic to be used in classical hypotheses tests.[8] Several coherent techniques have been hitherto proposed.[9–16]

Conversely, we investigate the potential of neural-network based classifiers[17,18] for discriminating GWBs of a given canonical family (e.g. supernovae (SNe) core-collapse waveforms) from typical transient instrumental artifacts, in the (noisy) data of a single detector. The further classification of glitches into typical sets is also explored.

In order to provide a proof of concept, we use the SNe core-collapse waveform catalog produced by Dimmelmeier and co-Workers,[19] and the data base of glitches observed in laser interferometer gravitational wave observatory (LIGO) data maintained by Saulson[20] to construct datasets of (windowed) transient waveforms (glitches and bursts) in additive (Gaussian and compound-Gaussian) noise with different signal-to-noise ratios (SNR). Principal component analysis (PCA)[21,22] is next implemented for reducing data dimensionality, yielding results consistent with, and extending those in the literature.[23]

Then, a multiLayer perceptron neural network is trained by a backpropagation algorithm (MLP-BP)[22,24] on a data subset, and used to classify the transients as glitch or burst.

A Self-Organizing Map (SOM) architecture[25–27] is finally used to classify glitches. The glitch/burst discrimination and glitch classification abilities are gauged in terms of the related truth tables.

There is a number of applications of neural networks for the analysis of the data collected by the new generation of instruments for astroparticle physics such as, for instance, the Solar energetic proton events[28] the cosmic ray telescopes AUGER[29,30] and ARGO[31,32]; the gamma ray Cherenkhov telescope,[33,34] the VIRGO GWs interferometer[35,36] and even for the search of the Higgs boson.[37] However, our results appear to be new.

This paper is organized as follows: In Sec. 2, we describe the data preprocessing and the structure of the neural network used for glitch/burst discrimination; results are presented in Sec. 2.4. The application of SOM to Glitch classification is discussed in Sec. 3 and the results are reported in Sec. 3.3. Concluding remarks are presented in Sec. 4.

## 2. Glitch-Burst Discrimination

A first-set of numerical experiments is aimed at exploring the possibility of training a neural network to discriminate between typical instrumental glitches and typical GWB of a specific (yet broad) family at a single detector level.

### 2.1. *Data set*

The glitch data set has been extracted from Saulson catalog[20] where a representative high SNR sub set of (noisy) glitches are stored; the files contain time domain (dimensionless) values of the detector output sampled at $16\,384\,\mathrm{Hz}$.

The files containing the clean bursts (simulated by means of numerical relativistic hydro-dynamical codes) are taken by Dimmelmeier and co-Workers.[19] The data-files in Ref. 19 are simulated waveform of SNe collapse unevenly sampled. We (polynomial) interpolated the bursts data and re-sampled the signals to the same sampling frequency of the glitches.

Finally, our basic data set consists of 68 GWB and 60 glitch waveforms of 150 ms, sampled at $16\,384\,\mathrm{Hz}$. All waveforms are preliminary passbanded and smoothed, so as to obtain fiducially noise-free prototypes. They are further re-scaled to unit $L_2$ norm, and window-centered before adding white (passband) Gaussian noise for a prescribed SNR (see Figs. 1 and 2).

### 2.2. *Principal component analysis*

PCA of the whole data set was needed in order to reduce the (highly redundant) dimension of the waveforms space. PCA is a common data analysis technique used to reduce the dimensionality of large data sets displaying a high degree of covariance. The original data set undergoes a linear transformation, whereby the axes of the new co-ordinate system are aligned such that the maximum variance is parallel to the first axis (principal component PC), the second highest variance lies along the second principal axis, etc. This is achieved by finding the principal eigenvalues of the co-variance matrix, $X$, and then applying criteria to the eigenvalues to optimize the dimensionality of the data space.

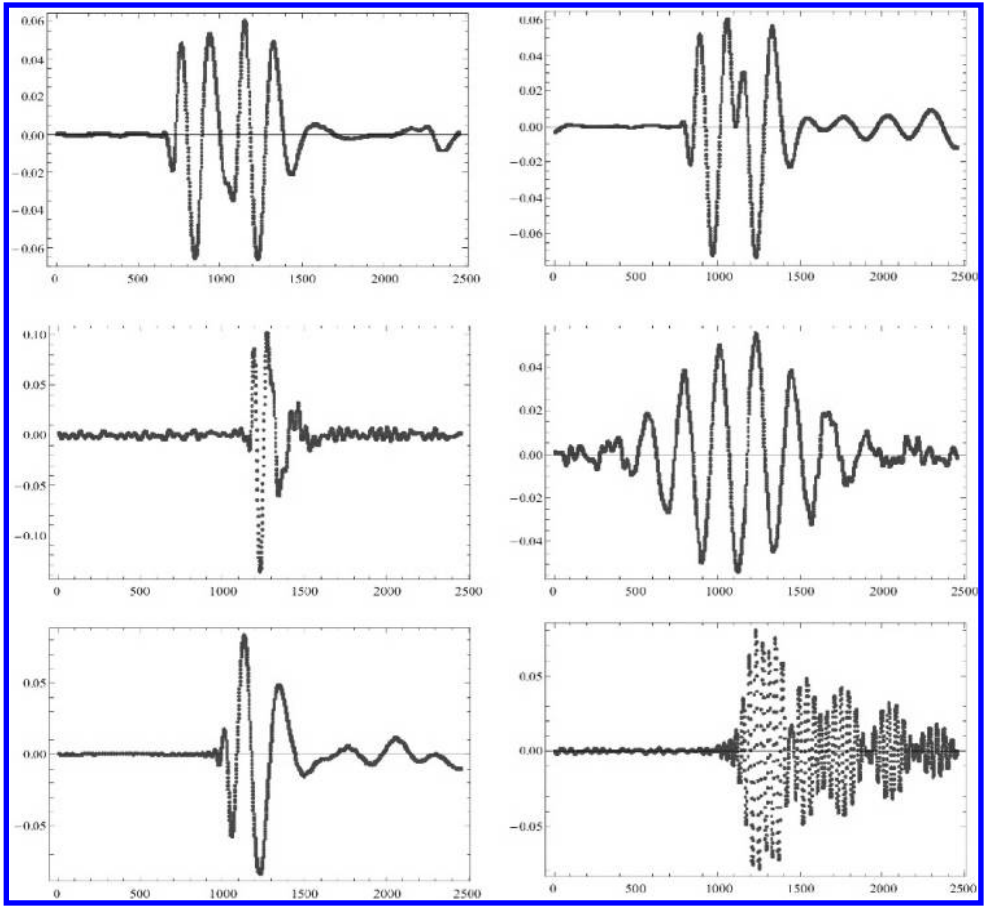

Fig. 1.   A few typical glitches in the dataset. In each figure, the *X*-axis represents the sample number and the *Y*-axis the corresponding amplitude.

For noise-free waveforms, it was found that 20 PCs accounted for almost 100% of the variance for the whole dataset (GWBs and glitches) (Fig. 21). For noisy data down to SNR $\sim 10$ the number of needed PCs was 35. We accordingly encoded all waveforms in the data set using 35-dimensional vectors of PCs.

This results extends to burst + glitch data the result by Heng[23] based on PCA, indicating that the effective dimension of the core-collapse supernova waveforms may be fairly small.

### 2.3.  *MLP-BP neural network*

Here, our goal is to setup a neural network capable of classify observed transients as GWBs or noise glitches. The network has been settled as a feedforward MLP-BP.

A feedforward MLP neural network is a computational structure made by many processing elements (units) — the neurons — operating in parallel.[38] These neurons are organized into clusters or layers. They are grouped in "input," "output" and "hidden" (i.e. those units which are neither input nor output) layers. Each neuron of a given layer is connected to all the neurons of the next one. The topological structure of the MLP is depicted in Fig. 4.

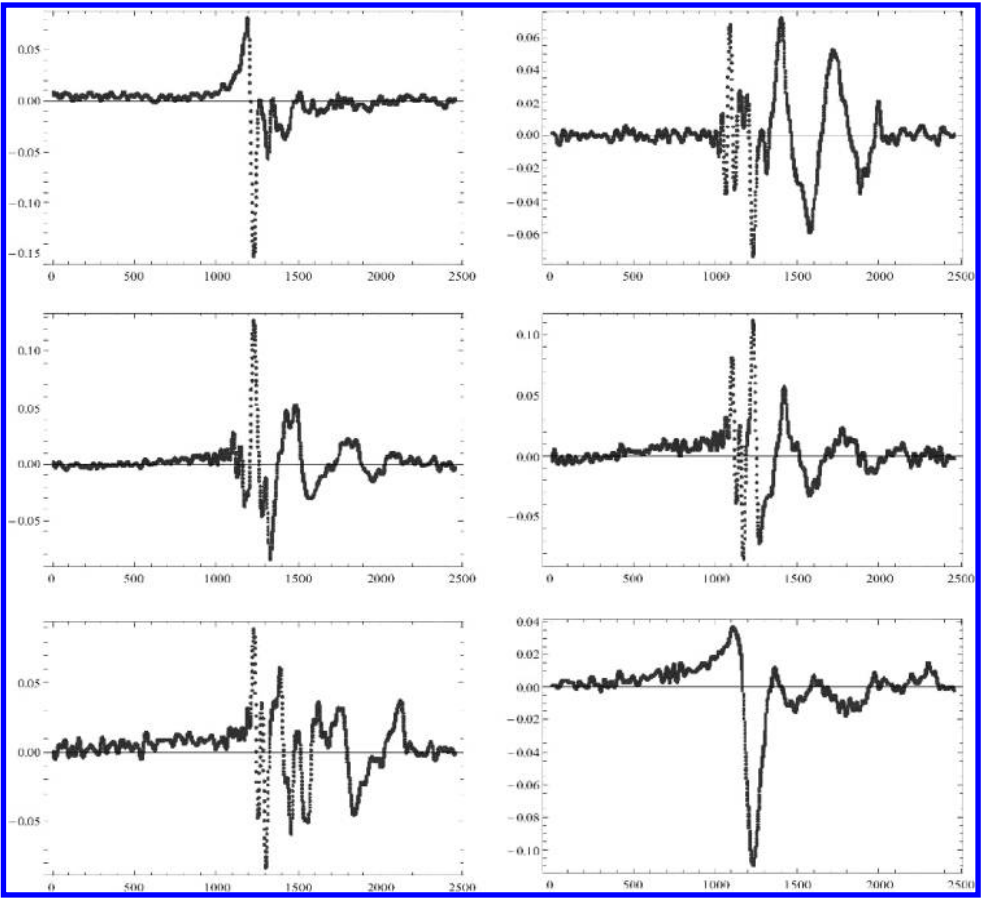

Fig. 2.   A few typical bursts in the dataset. In each figure, the *X*-axis represents the sample number and the *Y*-axis the corresponding amplitude.

Our model is synchronous: at each time every neuron receives as input the weighted sum of the input patterns and/or of the other neuron outputs, as shown in the following equation:

$$O_k = f\left[\sum_n W_{kn}O_n - B_k\right],\tag{1}$$

where $W_{kn}$ is the weight associated to the link from neuron $n$ to neuron $k$, $O_n$ is the output of neuron $n$ or of the $n$th input and $B_k$ is the neuron threshold, generally called bias. Then, the neuron output is a continuous and derivable function of its input with values comprised in the $[0, 1]$ range. For our experiments, the function $f$ is the sigmoid function

$$f(x) = 1/(1 + e^{-x}). \tag{2}$$

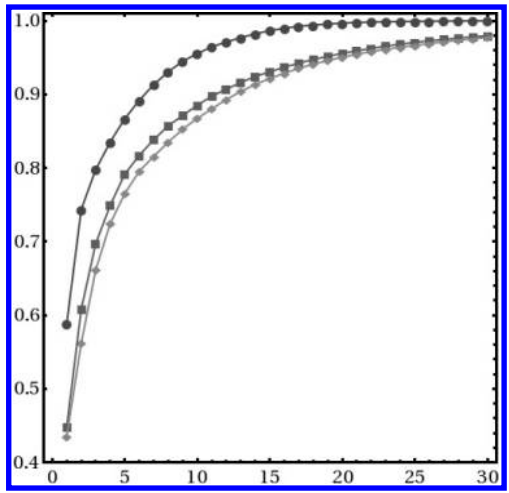

Fig. 3.    Results of the PCA for Clean (dots), SNR = 9 (square), and SNR = 7 (diamond) data. The *X*-axis reports the number of PCs and *Y*-axis the explained variance.

The training procedure is the so-called "back propagation."[39] It makes use of a subset of the data set feature vectors, each one labeled with the correct output, as examples of the correct input/output relationship. First, a vector is presented to the input neurons and then the network gives its output. If it is not equal to the desired one, the difference (error) between these two values is computed and the weights $W_{kn}$ are changed in order to minimize it. Given the $p$th pattern in input, the error $E_p$ is:

$$E_p = 1/2 \times \sum_j \left( t_{pj} - O_{pj} \right)^2, \tag{3}$$

where $t_{pj}$ is the $p$th desired output value and $O_{pj}$ is the output of the corresponding neuron.

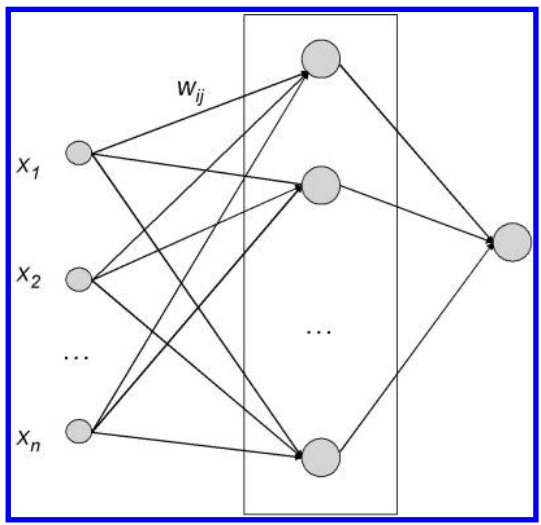

Fig. 4.    The topological structure of the MLP. The neurons are organized into clusters or layers: "input" (receiving the $x_1, x_2, \ldots, x_n$ vector), "output" (giving the classification value) and "hidden" (i.e. those units which are neither input nor output) layers. Each neuron of a given layer is connected to all the neurons of the next one by a weighted connection, $w_{ij}$.

These operations are repeated for all the vectors in the subset, so completing a so called *epoch* or *cycle*, and the process is iterated, until we minimize the Mean Square Error (MSE)[22] of the system.

## 2.4. *Results*

According to the 10-fold cross-validation methodology,[22] the whole data set has been partitioned into 10 non overlapping subsets. Of these 10 subsets, one can be chosen to evaluate the neural network performance (validation set), and the remaining 9 can be used to instruct it (training sets), i.e. setting the network weights. This procedure has been repeated 10 times, corresponding to all possible choices of the validation set and the corresponding average performance has been gauged in terms of an average %-misclassification error. Cross-validation is important in testing hypotheses suggested by the data, especially where further samples are costly or impossible to collect.

It was found by trial and error that two hidden layers were needed for successful operation with realistic noisy data, down to SNR $\sim 10$.

As shown in the left-hand side of Fig. 5, in the noise-free case, after 20 learning cycles (aka epochs), the misclassification error drops to a minimum, for both the training and validation sets, and remains almost constant as the number of cycles is further increased. In the noisy case (right-hand side of Fig. 5), the error drops to a minimum after 40 learning cycles, for both the training and validation sets. As the number of cycles is further increased, a transient blow up of the error is observed, followed by re-settling at the previously reached low level. This can be interpreted as due to temporary trapping of the neural network into a local minimum. These results help setting the number of neural network learning cycles in such a way that the misclassification error reaches a minimum both in the training phase (which indicates that the network learns at its best), and in the validation phase (which indicates that the network classifies at its best). On the basis of our simulations, we set the number of cycles $= 100$.

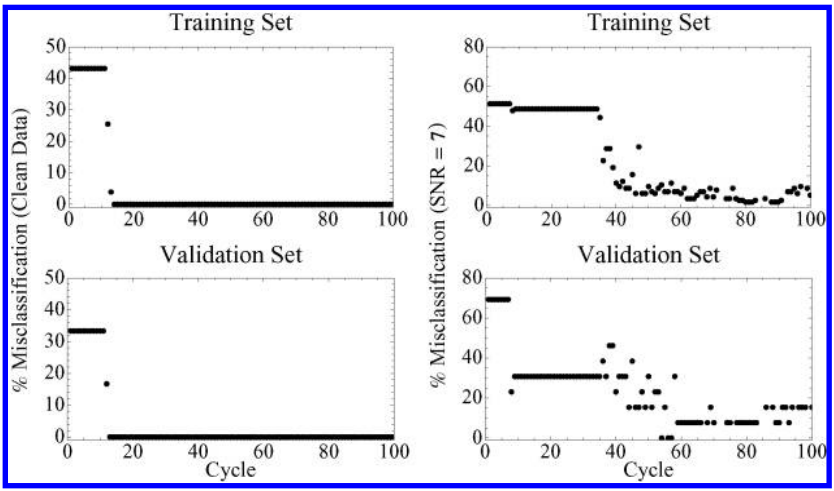

Fig. 5.    Training and validation set error behavior on clean (left-hand side) and noisy (right-hand side) data, growing the number of epochs.

The neural-network based glitch-burst discriminator works uniformly well (with an average misclassification error $< 5\%$) down to SNR $\sim 10$. Remarkably, all observed misclassification errors are conservative (a few GWBs are misclassified as glitches — false dismissal, but not vice versa — false alarm). For SNR $< 6$, the network fails almost completely in discriminating GWB from glitches.

As shown in Table 1, when there is no noise in the data we reach a misclassification error of 4.62%.

Table 2 shows the results obtained on SNR $= 9$ data, by varying the number of (PC). The table reports for each test the number of errors (nE) and the error percentage (%), and, for each group of ten trials, the mean and the Standard Deviation (StD). The best results are obtained by using 35 PCs, reaching a misclassification error of 6.25%.

Table 3 shows the results by using a SNR equal to 7, varying the number of PCs. Also in this case, the error is generally low and reaches the minimum (mean error 7.81%) on 35 features.

By further lowering to 6, the SNR ability to distinguish between glitch and burst signals is lost. The error grows to 50% (random classification). Table 4 resumes the most significant experiments.

Table 1. 10-fold cross-validation results by using the clean data (100 epochs).

| Test | % |
|------|-------|
| 1 | 0.00 |
| 2 | 7.69 |
| 3 | 0.00 |
| 4 | 15.38 |
| 5 | 7.69 |
| 6 | 0.00 |
| 7 | 0.00 |
| 8 | 7.69 |
| 9 | 7.69 |
| 10 | 0.00 |
| Mean | 4.62 |

Table 2.   10-fold cross-validation results by using a SNR equal to 9.

| PC | 45 | | 40 | | 35 | | 30 | | 25 | |
|------|------|------|------|------|------|------|------|------|------|------|
| Test | nE | % | nE | % | nE | % | nE | % | nE | % |
| 1 | 2 | 15.63 | 1 | 7.81 | 0 | 0.00 | 2 | 15.63 | 0 | 0.00 |
| 2 | 1 | 7.81 | 2 | 15.63 | 1 | 7.81 | 2 | 15.63 | 1 | 7.81 |
| 3 | 2 | 15.63 | 2 | 15.63 | 1 | 7.81 | 0 | 0.00 | 2 | 15.63 |
| 4 | 0 | 0.00 | 0 | 0.00 | 1 | 7.81 | 3 | 23.44 | 1 | 7.81 |
| 5 | 1 | 7.81 | 1 | 7.81 | 2 | 15.63 | 0 | 0.00 | 1 | 7.81 |
| 6 | 1 | 7.81 | 2 | 15.63 | 2 | 15.63 | 1 | 7.81 | 0 | 0.00 |
| 7 | 2 | 15.63 | 1 | 7.81 | 0 | 0.00 | 2 | 15.63 | 0 | 0.00 |
| 8 | 0 | 0.00 | 2 | 15.63 | 0 | 0.00 | 0 | 0.00 | 4 | 31.25 |
| 9 | 2 | 15.63 | 2 | 15.63 | 1 | 7.81 | 1 | 7.81 | 2 | 15.63 |
| 10 | 0 | 0.00 | 0 | 0.00 | 0 | 0.00 | 3 | 23.44 | 1 | 7.81 |
| Mean | 1.1 | **8.59** | 1.3 | **10.16** | 0.8 | **6.25** | 1.4 | **10.94** | 1.2 | **9.38** |
| StD | | **0.88** | | **0.82** | | **0.79** | | **1.17** | | **1.23** |

Table 3.   10-fold cross-validation results by using a SNR equal to 7.

| PC | 50 | | 40 | | 35 | |
|------|------|------|------|------|------|------|
| Test | nE | % | nE | % | nE | % |
| 1 | 1 | 7.81 | 0 | 0.00 | 0 | 0.00 |
| 2 | 1 | 7.81 | 1 | 7.81 | 1 | 7.81 |
| 3 | 2 | 15.63 | 3 | 23.44 | 2 | 15.63 |
| 4 | 4 | 31.25 | 1 | 7.81 | 1 | 7.81 |
| 5 | 3 | 23.44 | 0 | 0.00 | 1 | 7.81 |
| 6 | 3 | 23.44 | 2 | 15.63 | 1 | 7.81 |
| 7 | 3 | 23.44 | 3 | 23.44 | 1 | 7.81 |
| 8 | 2 | 15.63 | 0 | 0.00 | 1 | 7.81 |
| 9 | 3 | 23.44 | 1 | 7.81 | 1 | 7.81 |
| 10 | 4 | 31.25 | 1 | 7.81 | 1 | 7.81 |
| Mean | 2.6 | **20.31** | 1.2 | **9.38** | 1 | **7.81** |
| StD | | **1.07** | | **1.14** | | **0.47** |

Table 4.   10-fold cross-validation results by varying the SNR.

| SNR | Hidden layers | Cycles | PC | 10-fold cross-validation |
|------|------|------|------|------|
| — | 2 | 100 | 25 | 4.62% |
| 9 | 2 | 100 | 35 | 6.25% |
| 7 | 2 | 100 | 35 | 7.81% |
| 6 | 2 | 100 | 35 | 50% |

The neural network results compare favorably with a state of art method[40] discriminating between GW and glitches by using Cross Wigner Spectra, in term of misclassification error (false alarm and false dismissal). As illustrated in Fig. 6, the Cross Wigner Spectra method reaches the neural network performances at a much higher SNR ratio.

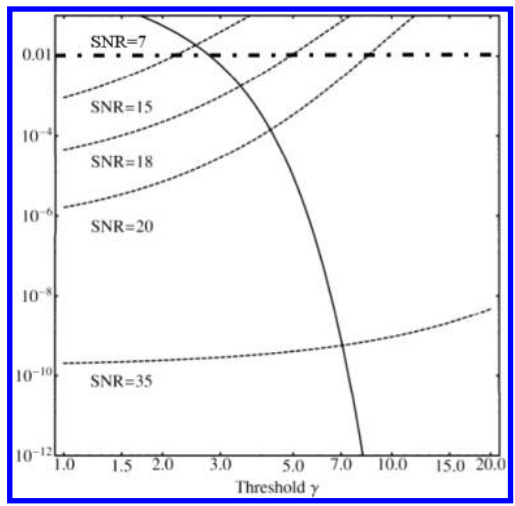

Fig. 6. The false alarm (continuous line) and false dismissal (the dashed lines) probabilities, corresponding to different values of the SNR as functions of the detection threshold $\gamma$ for the Cross Wigner Spectra method, and the corresponding performances of the neural network (long–short dashes) by using a SNR equal to 7.

Figure 6 displays the false alarm (continuous line) and false dismissal probabilities (the dashed lines), corresponding to different values of the SNR in the 15–35 range as functions of the detection threshold $\gamma$, a free parameter by using Cross Wigner Spectra, and the corresponding performances of the neural network (long–short dashes) by using a SNR equal to 7, where the detection threshold is implicitly fixed by the learned threshold $B_k$ (1) of the neural network output neuron.

## 3. Glitch Classification

Here, we perform unsupervised clustering of instrumental/environmental glitches. The purpose is to identify and characterize clusters in dataset of glitches. Identification of subgroups within the database could help to elucidate signal characteristics and facilitate future model building, also taking into the account the interferometer architecture.

### 3.1. *Data set*

We add to the 60 glitches from the Saulson catalog[20] the subset of 27 "labeled" glitches in Ref. 41, reported in Fig. 7 for which a physical origin had been unambiguously traced out.

For homogeneity, we used the same preprocessing of Sec. 2.1. Finally our basic data set consists of 87 glitch waveforms of 150 ms, sampled at 16 384 Hz. The application of PCA to the glitch data led to a dramatic dimension reduction to 15 components, accounting for over the 95% of the variance.

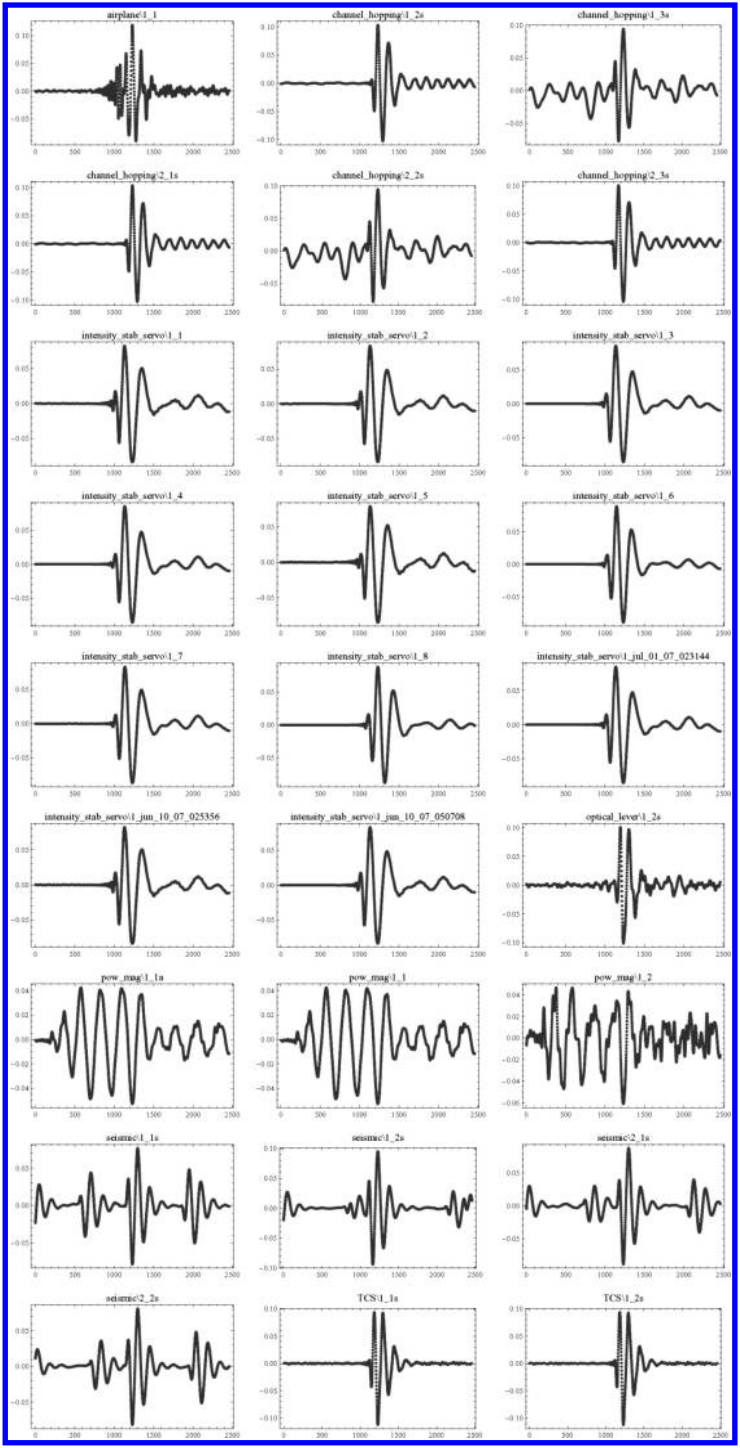

Fig. 7. "Labeled" glitches.

## 3.2. *Self-organizing map*

The glitch clustering is obtained by means of Kohonen SOM. SOMs is a class of artificial neural networks aimed at mapping a set of input data to few clusters that serve as prototypes. Differently from other neural networks, a SOM is able to preserve the topological properties of the input data, thus allowing to visualize the relationship between the groups.

A SOM is structured as single layer feed-forward neural network, in which each neuron in connected to the others in order to form a lattice, as depicted in Fig. 8.

The network is trained using an unsupervised learning algorithm. The goal of learning in a SOM is to make the network responding, similarly to certain input patterns. The initial setting of neuron weights can be fully random or driven by the two largest PC eigenvectors, in order to speed-up the learning process as the initial weights already give good approximation of SOM weights.

The training procedure is based on competitive learning, organized in the following steps:

(1) An input vector $\mathbf{x}(t)$ is compared with all the weight vectors (*centers*) $\mathbf{w}_i(t)$. The best-matching unit (BMU) on the map, i.e. the neuron where the weight vector is closest to the input vector according to some metric (e.g. Euclidean), is identified.
(2) The weight vectors of the winner and a number of its neighboring neurons in the array are changed towards the input vector.

The adaptation of the model vectors in the learning is driven by equations:

$$\mathbf{w}_i(t+1) = \begin{cases} \mathbf{w}_i(t) + \alpha(t)[\mathbf{x}(t) - \mathbf{w}_i(t)], & i \in N_c(t), \\ \mathbf{w}_i(t), & \text{otherwise,} \end{cases} \tag{4}$$

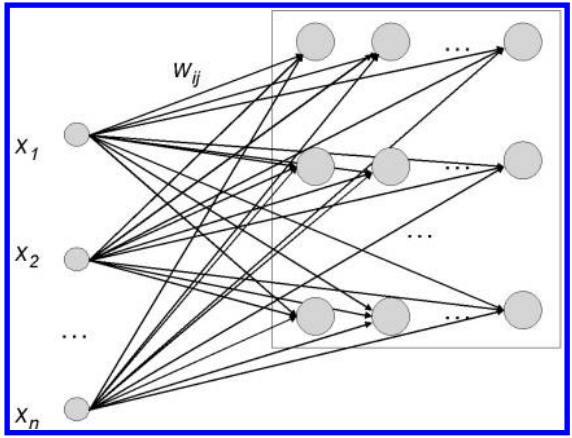

Fig. 8.   The topological structure of the Kohonen SOM.

where $t$ is the discrete-time index, $\alpha(t) \in [0,1]$ is a scalar defining the *learning rate*, and $N_c(t)$ specifies the neighborhood around the BMU in the map lattice. In the simplest form, $N_c(t) = 1$ for all neurons close within a range to BMU and $N_c(t) = 1$ for others. Here, we use another functional form (e.g. Gaussian) for $N_c$, and the neighborhood shrinks over the time,[42] being broader at the beginning and smaller when converging to local estimates. The factor $\alpha(t)$ also decreases during learning.

This learning process is iterated over a (usually large) number of cycles, thus requiring to present re-iteratively the input sample to the SOM algorithm. During the learning process, individual changes may be contradictory, but convergence towards ordered values for the $\mathbf{w}_i(t)$ emerge along the process.[42–45]

The network learning ends associating output nodes with groups of patterns in the input data set.

### 3.3. *Results*

We used a SOM based on the nearest-neighbor topology with Euclidean distance. After 1500 cycles, the number of clusters found settled to 8, plus a single "orphan." Remarkably, this gave the lowest (best) value of the Davies–Bouldin cluster separation measure[46] among all alternative unsupervised classification yielding a different (fixed) number of clusters. Clusters can be represented by a central vector (*centroid*), which may not necessarily be a member of the data set. Our cluster centroids are depicted in Fig. 9, while the relative distances between the centers and from the overall mean are reported in Table 5 and Fig. 10, respectively.

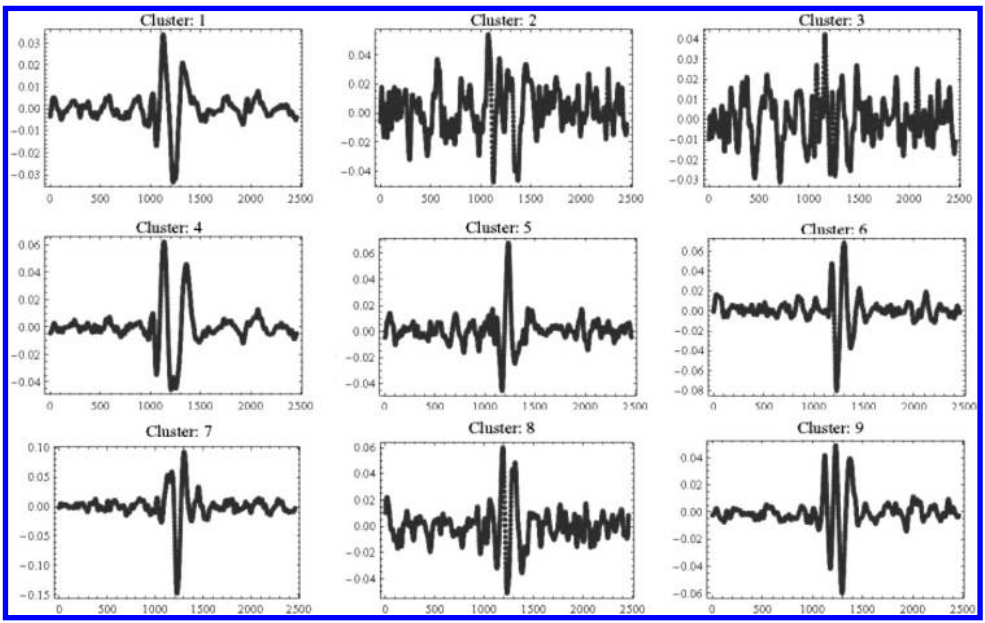

Fig. 9. Cluster centroids.

Table 5.  Relative distances between the centers.

| Cluster | Overall | 1 | 2 | 3 | 4 | 5 | 6 | 7 | 8 | 9 |
|---------|---------|---|---|---|---|---|---|---|---|---|
| Overall | **0.00** | 0.94 | 0.58 | 0.95 | 0.41 | 0.80 | 0.72 | 0.91 | 0.74 | 0.73 |
| 1 | 0.94 | **0.00** | 0.88 | 1.35 | 1.27 | 1.00 | 0.93 | 1.33 | 0.75 | 1.22 |
| 2 | 0.58 | 0.88 | **0.00** | 1.18 | 0.82 | 0.89 | 0.92 | 1.14 | 0.76 | 0.91 |
| 3 | 0.95 | 1.35 | 1.18 | **0.00** | 1.07 | 1.67 | 0.75 | 0.20 | 1.03 | 1.67 |
| 4 | 0.41 | 1.27 | 0.82 | 1.07 | **0.00** | 1.04 | 1.06 | 1.03 | 1.11 | 0.77 |
| 5 | 0.80 | 1.00 | 0.89 | 1.67 | 1.04 | **0.00** | 1.17 | 1.63 | 1.04 | 0.56 |
| 6 | 0.72 | 0.93 | 0.92 | 0.75 | 1.06 | 1.17 | **0.00** | 0.70 | 0.55 | 1.35 |
| 7 | 0.91 | 1.33 | 1.14 | 0.20 | 1.03 | 1.63 | 0.70 | **0.00** | 1.01 | 1.63 |
| 8 | 0.74 | 0.75 | 0.76 | 1.03 | 1.11 | 1.04 | 0.55 | 1.01 | **0.00** | 1.21 |
| 9 | 0.73 | 1.22 | 0.91 | 1.67 | 0.77 | 0.56 | 1.35 | 1.63 | 1.21 | **0.00** |

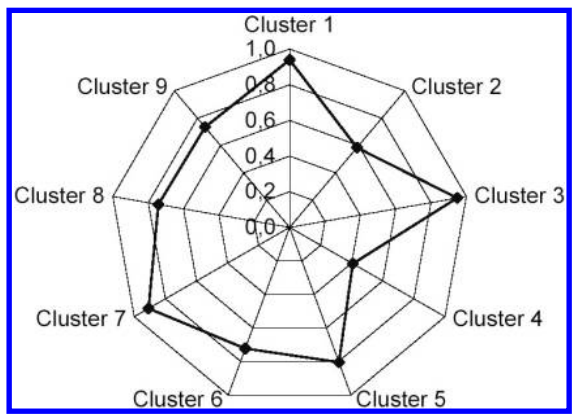

Fig. 10.  Distances of each cluster center from the overall mean.

Table 6.  Glitches for which a physical origin had been traced out, fitted into the obtained SOM-determined clusters.

| Origin | Cluster |
|--------|---------|
| $airplane\backslash 1\_1.flt$ | 5 |
| $channel\_hopping\backslash 1\_2s.flt$ | 9 |
| $channel\_hopping\backslash 1\_3s.flt$ | 9 |
| $channel\_hopping\backslash 2\_1s.flt$ | 9 |
| $channel\_hopping\backslash 2\_2s.flt$ | 9 |
| $channel\_hopping\backslash 2\_3s.flt$ | 8 |
| $intensity\_stab\_servo\backslash 1\_1.flt$ | 4 |
| $intensity\_stab\_servo\backslash 1\_2.flt$ | 4 |
| $intensity\_stab\_servo\backslash 1\_3.flt$ | 4 |
| $intensity\_stab\_servo\backslash 1\_4.flt$ | 4 |
| $intensity\_stab\_servo\backslash 1\_5.flt$ | 4 |
| $intensity\_stab\_servo\backslash 1\_6.flt$ | 4 |
| $intensity\_stab\_servo\backslash 1\_7.flt$ | 4 |
| $intensity\_stab\_servo\backslash 1\_8.flt$ | 5 |
| $intensity\_stab\_servo\backslash 1\_jul\_01\_07\_023144.flt$ | 4 |

Table 6. (*Continued*)

| Origin | Cluster |
|---|---|
| $intensity\_stab\_servo \backslash 1\_jun\_10\_07\_025356.flt$ | 4 |
| $intensity\_stab\_servo \backslash 1\_jun\_10\_07\_050708.flt$ | 4 |
| $optical\_lever \backslash 1\_2s.flt$ | 8 |
| $pow\_mag \backslash 1\_1.flt$ | 9 |
| $pow\_mag \backslash 1\_1a.flt$ | 4 |
| $pow\_mag \backslash 1\_2.flt$ | 2 |
| $seismic \backslash 1\_1s.flt$ | 6 |
| $seismic \backslash 1\_2s.flt$ | 9 |
| $seismic \backslash 2\_1s.flt$ | 6 |
| $seismic \backslash 2\_2s.flt$ | 6 |
| $TCS \backslash 1\_1s.flt$ | 6 |
| $TCS \backslash 1\_2s.flt$ | 3 |

Table 7.   Cluster sizes.

| Cluster | 1 | 2 | 3 | 4 | 5 | 6 | 7 | 8 | 9 |
|---|---|---|---|---|---|---|---|---|---|
| Size | 1 | 3 | 14 | 26 | 11 | 7 | 4 | 4 | 17 |

It was tempting to check how/whether the subset of 27 "labeled" glitches for which a physical origin had been unambiguously traced out, fitted into the obtained SOM-determined clusters. The results of this check are summarized in the Table 6. The majority (10 out of 11) of $int\_stab\_servo$ labeled glitches fell into cluster 4, which comprises 26 elements (see also Table 7); the majority of $ch\_hopping$ labeled glitches (4 out of 5) fell into cluster 9, which comprises 17 elements and the majority of seismic labeled glitches (3 out of 4) fell into cluster #6, which comprises 7 elements. Indeed, the waveforms corresponding to the centroids of clusters #4, #9 and #6 are suggestively similar to those corresponding to the labeled glitch subsets $int\_stab\_servo$, $ch\_hopping$ and $seismic$ from Ref. 41.

## 4. Concluding Remarks

A two-hidden layer MLP neural network engine performs reasonably well in discriminating glitches from GWBs, for the specific GWB and glitch families used here, down to SNR values of the order of 10 — a result not obvious *a priori*. Also, a neural-network based SOM classifier produced a relatively small number of clusters out of the glitch set, some of which accommodate selectively the majority of specific "labeled" glitches of known physical origin. More extensive work is obviously needed in order to validate/sharpen these results, and make them useful for the experiments.

While, some works[47] suggests that single-detector trigger classification is a daunting task even with auxiliary channel information, our preliminary results, confirmed by a 10-fold cross-validation methodology, suggest that the approach is effective and robust throughout the SNR range of practical interest. We evidenced the potential of neural-network based classifiers for discriminating GWBs in the data of a single detector.

Versatility (robustness to signal model uncertainty) and algorithmic simplicity are two advantages of this methodology when compared to more standard approaches for the detection of such waveforms.

Perspective applications pertain both to distributed (network, multisensor) detection of GWBs, where some *intelligence* at the single node level can be introduced, and instrument diagnostics/optimization, where spurious transients can be identified, classified and hopefully traced back to their entry points.[48]

## Acknowledgment

One author (S.R.) wishes to thank the colleague and friend A. Feoli for the useful discussions.